# 2-10 μm Mid-infrared Supercontinuum Generation in As$_2$Se$_3$ Photonic Crystal Fiber

**Wu Yuan**

*Singapore Institute of Manufacturing Technology, 71 Nanyang Drive, Singapore 638075*
*wyuan@simtech.a-star.edu.sg*

**Abstract:** For the first time, we demonstrated that a hyper-broadband from 2μm to 10μm can be generated with a high spectral flatness by using a ~4 um pump and an As$_2$Se$_3$ photonic crystal fiber. The broad and flat dispersion profile and the low guiding loss of the endlessly single-mode As$_2$Se$_3$ photonic crystal fibers combining with the high nonlinearity and wide mid-infrared transparent window of As$_2$Se$_3$ chalcogenide glass are essential for this high performance mid-infrared laser source.

## 1. Introduction

Supercontinuum (SC) generation in optical fiber is a complex process of interplaying of Kerr effect, Raman effect, and dispersion properties of fiber, it has been widely demonstrated and well understood in all pump regimes ranging from continuous wave, nanosecond, picosecond to ultrashort femtosecond [1]. Commercial SC sources are now available using primarily lasers at 1064 nm and can provide a bandwidth from 400–2400 nm. The fiber of choice has traditionally been the silica PCF because it can be made endlessly single-mode, and its dispersion can be tailored and its zero-dispersion wavelength (ZDW) can move down into the visible region. The intense research interest on SC generation now has been kept and extended further into the mid-infrared (MIR) spectral region in recent years due to its potential usefulness in a variety of applications requiring broad MIR spectrum [2-9].

A great variety of soft-glass fibers and pump lasers have been considered to develop the broadband MIR SC sources[3-9]. Using soft-glass fibers such as fluoride, chalcogenide, and tellurite fibers for MIR SC generation is due to their higher nonlinearity and lower transmission loss in the MIR regime comparing with these of silica fibers, which are not transparent for wavelength beyond 2.5 μm. Compared to fluroide and tellurite glass, chalcogenide glass, in particular As$_2$Se$_3$ glass, has a higher nonlinear index. Moreover, the long-wavelength side of the transmission windows of tellurite and fluride glass are shorter than 5 μm [3,10], which sets the long-wavelength limit of SC generation to at best 4.5 μm [9]. By contrast, the transmission window of As$_2$Se$_3$ glass cuts off at about 10 μm at the long-wavelength side [10,11], it potentially allows the generation of significant supercontinuum radiation beyond 6μm, which is high desirable for the varaint advanced MIR spectrascopical applications. However, As$_2$Se$_3$ glass has a larger refractive index than both fluriode and tellurite glass, it is difficult to find a pump source to work around the ZDW of As$_2$Se$_3$ at about 5 μm. Employing the PCF design is a valid method to tailor the dispersion of As$_2$Se$_3$ fiber and move the ZDW down to a shorter wavelength.

As the reliable pump light sources in 1.8 to 2.1 μm wavelength regime, mode-locked thulium fiber lasers have attracted increasing interest [7-9]. Recently, a new kind of broadband MIR SC source with bandwith from 2 to 4.6 μm has been demonstrated by pumping a tapered chalcogenide fiber with a subharmonic generation source of the mode-locked Erbrium fiber laser from a degenerated optical parametric oscillator (OPO) [12]. This new pump scheme is very compact and useful, it leads to a highly potential solution of an efficient pump source for generating MIR wavelength beyond 6um with As$_2$Se$_3$ fiber. Based on this, we perform a realistic investigation of obtaining the ultrabroad MIR SC generation in a relatively short length of highly nonlinear As$_2$Se$_3$ PCF fiber by using a subharmonic generation source of the mode-locked thulium fiber laser at ~4.1 μm [13].

## 2. Fiber design and SC generation model

Here we consider the simple five ring triangular PCF structures, which can be fabricated by utilizing standard extrusion and stacking based methods. While keeping the relative hole size (d/ Λ = 0.4) constant to assure that the fibers are endlessly single-mode [14], the pitch of PCF is varied (Λ= 3, 4, 5, and 6 μm). As indicated by the material loss measurements of Naval Research Laboratory, it turns out that the generated SC will be severely limited by the material absorption at < ~2 μm and > ~9 μm [3,15].

For the refractive index of As$_2$Se$_3$, an experimentally derived Sellmeier equation is proposed by Thompson of Amorphous Material Inc. at 2008 [16]:

$$n(\lambda) = \left[1 + \lambda^2[A_0^2/(\lambda^2 - A_1^2) + A_2^2/(\lambda^2 - 19^2) + A_3^2/(\lambda^2 - 4 \times A_1^2)]\right]^{0.5} \quad [1]$$

where λ is the wavelength, A$_0$, A$_1$, A$_2$, A$_3$ are Sellmeier coefficients and A$_0$=2.234921, A$_1$=0.24164, A$_2$=0.347441, A$_3$=1.308575, respectively. The refractive index calculated with Thompson's Sellmeier equation is compared with the Sellmeier equation of Cheirf et al. [17]:

$$n(\lambda) = [A + B\lambda^2/(\lambda^2 - D) + C\lambda^2]^{0.5} \qquad [2]$$

where λ is the wavelength, A = -4.5102, B = 12.0582, C = 0.0018, D = 0.0878. As show in Fig. 1, we find very good approximation between two refractive index models at near infrared range, in particualr, around 1.5 μm. Discrepancy is found and becomes biger when the wavelength is getting longer than 1.8 μm. Due to the spectral range of interest in MIR, it is therefore Thompson's Sellmeier equation of $As_2Se_3$ is prefered and employed in our calculations thereafter.

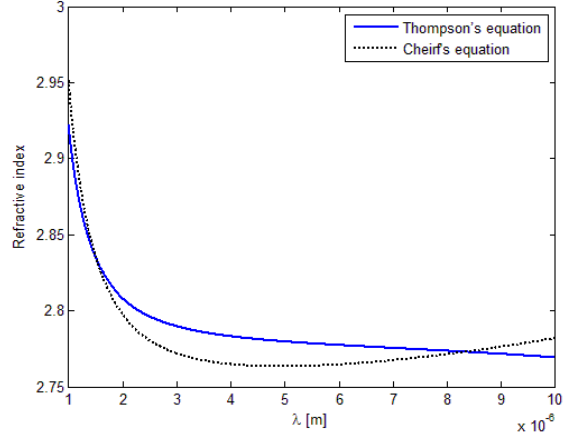

Fig.1.The calculated refractive index of $As_2Se_3$ by using different Sellmeier equations, blue solid line: Thompson's equation [16], black dotted line: Cheirf's equation [17].

By using the finite-element method based full-vector solver COMSOL, we calculate the effective index of the fundamental mode and its corresponding confinement loss of $As_2Se_3$ PCF fiber at the frequency of interest for different pitches. The resulting confiment losses of different pitches are shown in Fig 2 (a). For the fiber structure with Λ ≤ 3 μm, the confinement loss is significant at long wavelengths above ~7 μm, whereas in other cases confinement losses are negligible in comparison with material losses [3,15]. As it turns out, in order to achieve the maixaml SC bandwidth limited only by the intrinsic material loss, the $As_2Se_3$ PCF shall has a pitch of more than 3 μm to avoid the negative impacts of the confiment loss. In Fig. 2 (b), we plot the calculated dispersion for the $As_2Se_3$ PCF of different pitches. We find that, by choosing the pitch, a flat dispersion profile with an absolute dispersion within 30 ps/(nm km) can be achieved for spectral range from about 3 μm to 10 μm, with a double-ZDW for some pitch conditions. A low and flat dispersion profile is important for the maximization of the SC generation in fiber as it can decrease the temperal walk-off effect during the spectral boarding process. Note that the pumping wavelength of 4.1 μm is in the anomalous dispersion region for Λ = 3 and 4 μm, whereas the pulse experiences normal dispersion when Λ = 5 and 6 μm.

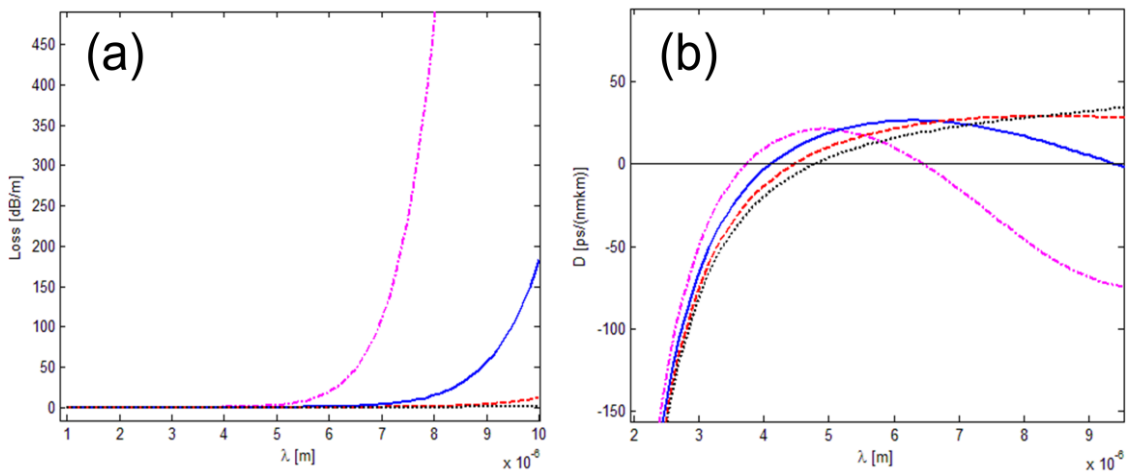

Fig.2. (a), The confinement loss, and (b) the dispersions of $As_2Se_3$ PCF with different pitches and a constant hole to pitch ratio of 0.4, Magenta dash dot line: pitch 3 μm, blue solid line: pitch 4 μm, red dashed line: pitch 5 μm, black dotted line: pitch 6 μm.

To study the pulse propagation and the SC generation, we employ the generalized nonlinear Schrödinger equation (GNLSE) in which the loss, higher order dispersion, stimulated Raman scattering, and frequency dependence of the nonlinear response is considered. We use a change in variables to shift into the so-called interaction picture in order to avoid the stiff dispersive part of the GNLSE and obtain the ordinary differential envelope equation [18,19]:

$$\frac{\partial \widetilde{A'}}{\partial Z} = i\bar{\gamma}(\omega)\exp(-i\hat{L}(\omega)Z) \times \mathcal{F}\{A(Z,T)\int_{-\infty}^{\infty} R(T')|A(Z,T-T')|^2 dT'\} \quad [3]$$

where the nonlinear response $\bar{\gamma}(\omega)$ is given by

$$\bar{\gamma}(\omega) = \frac{n_2 n_0 \omega}{c n_{eff}(\omega) A_{eff}^{1/4}(\omega)}$$

And

$$A(Z,T) = \mathcal{F}^{-1}\left\{\frac{\tilde{A}(Z,\omega)}{A_{eff}^{1/4}(\omega)}\right\}$$

Here $\mathcal{F}$ denotes Fourier transformation. The change in variables is given by $\widetilde{A'}(Z,\omega) = \tilde{A}(Z,\omega)\exp(-i\hat{L}(\omega)Z)$, where the linear operator $\hat{L}(\omega) = \beta(\omega) - \beta(\omega_0) - \beta_1(\omega_0)[\omega - \omega_0] + i\alpha(\omega)/2$.

To determine the Raman response function $R(t) = (1 - f_R)\delta(t) + f_R h_R(t)$, we derive $h_R(t)$ from the Raman gain spectrum measured by Naval Research Laboratory [15]. We have found that the the measured gain spectrum can be modeled by a two decaying harmonic oscillators function with the fractional contribution of $f_a$ and $f_b$ respectively:

$$h_R(t) = f_a \tau_1(\tau_1^{-2} + \tau_2^{-2})\exp(-\tau/\tau_2)\sin(\tau/\tau_1) + f_b \tau_3(\tau_3^{-2} + \tau_4^{-2})\exp(-\tau/\tau_4)\sin(\tau/\tau_3) \quad [4]$$

By choosing $f_a = 0.7$, $f_b = 0.3$, $\tau_1 = 23 \times 10^{-15}$, $\tau_2 = 230 \times 10^{-15}$, $\tau_3 = 20.5 \times 10^{-15}$, $\tau_4 = 260 \times 10^{-15}$, our model fits the experimental results very well with the main peak of the Raman gain located at $\Omega_R$ ~6.9 THz (~230 cm$^{-1}$), as shown in Fig. 3.

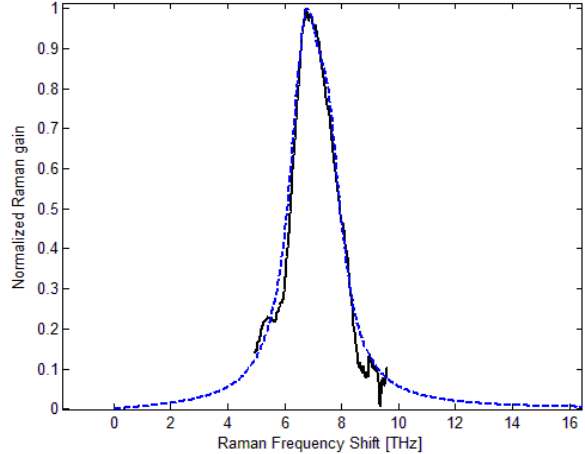

Fig.3. The Raman gain spectra of As$_2$Se$_3$, blue dashed curve: two decaying harmonic oscillators function model, black solid curve: measurement data of ref 15.

The nonlinear index n$_2$ of chalcogenides can be estimated by using the explicit Kramers-Kronig transformation equation which manifests the Kerr coefficient with linear index, Kerr coefficient dispersion, and glass parameters, as shown by Lenz et al.[20, 21]:

$$n_2(\nu) = 1.7 \times 10^{-18}(n^2 + 2)^3(n^2 - 1)\left(\frac{\sigma}{n \times E_s}\right)^2 F\left(\frac{h\nu}{E_g}\right) \quad [5]$$

where n$_2$ is in units of m$^2$/W, for As$_2$Se$_3$, $\sigma \approx 0.243 nm$, which is the mean anion-cation bond length of the bonds that are primarily responsible for the nonlinear response, Sellmeier gap $E_s = 4.1\ eV$, n is the linear refrative index estimated from Sellmeier equation of Eq.1, $F\left(\frac{h\nu}{E_g}\right)$ is dispersion function of Kerr coefficient which is proportional to the ratio of photon energy $h\nu$ and band gap energy $E_g$, $E_g$ of As$_2$Se$_3$ is $1.78\ eV$. For pump wavelength 4.1 µm, $F\left(\frac{h\nu}{E_g}\right)$ is about 1.2 [20]. Using Eq. 5 and Eq.1, we find that n$_2$= 5.74 ×10$^{-18}$ m$^2$/W at wavelength of 4.1 µm. This value is approximately 220 times the nonlinear index of silica at wavelength 1.5 µm, which is ~ 2.6 ×10$^{-20}$ m$^2$/W. We use f$_R$ = 0.1 as reported by Slusher et al [15]. The two photon absorption coefficient is neglected because the interested spectral region is far away from the main peak of two photon absorption at 1.39 µm [15].

We consider photonic crystal fibers (PCFs) with different Λ and in all cases we pump with a subharmonic generation source of the mode-locked thulium fiber laser which has a wavelength at 4.1 μm, a pulse duration of 500 fs and peak power of 10 kW. Noise is included by adding to the input pulse one random phase photon per mode in the frequency domain.

**3. Results and discussion**

The spectral and temporal evolutions of SC generation in the $As_2Se_3$ PCF with different pitches, i.e., Λ= 3, 4, 5, 6, and 7 μm, are shown in Fig. 4, with the resulting spectra of SC after the propagation of a 6 cm-long fiber. The dynamics of the SC generation when pumping in the spectral regime around ZDWs with femtosecond pulses is well known [1]. As we observe the formation of a symmetric sideband structure after only a few millimeters from Fig. 4 (a), when pumping in the anomalous regime for Λ= 3, 4 μm, the initial broadening arises due to four-wave mixing (FWM); when pumping in the normal dispersion regime for Λ= 5, 6, and 7 μm, the initial symmetric spectral broadening occurs mainly due to the higher-order dispersion phase-matched FWM [7]. The spectral broadening accompanied by a strong temporal compression at the intial stages of propagation is also noted, as shown in Fig. 4 (c).

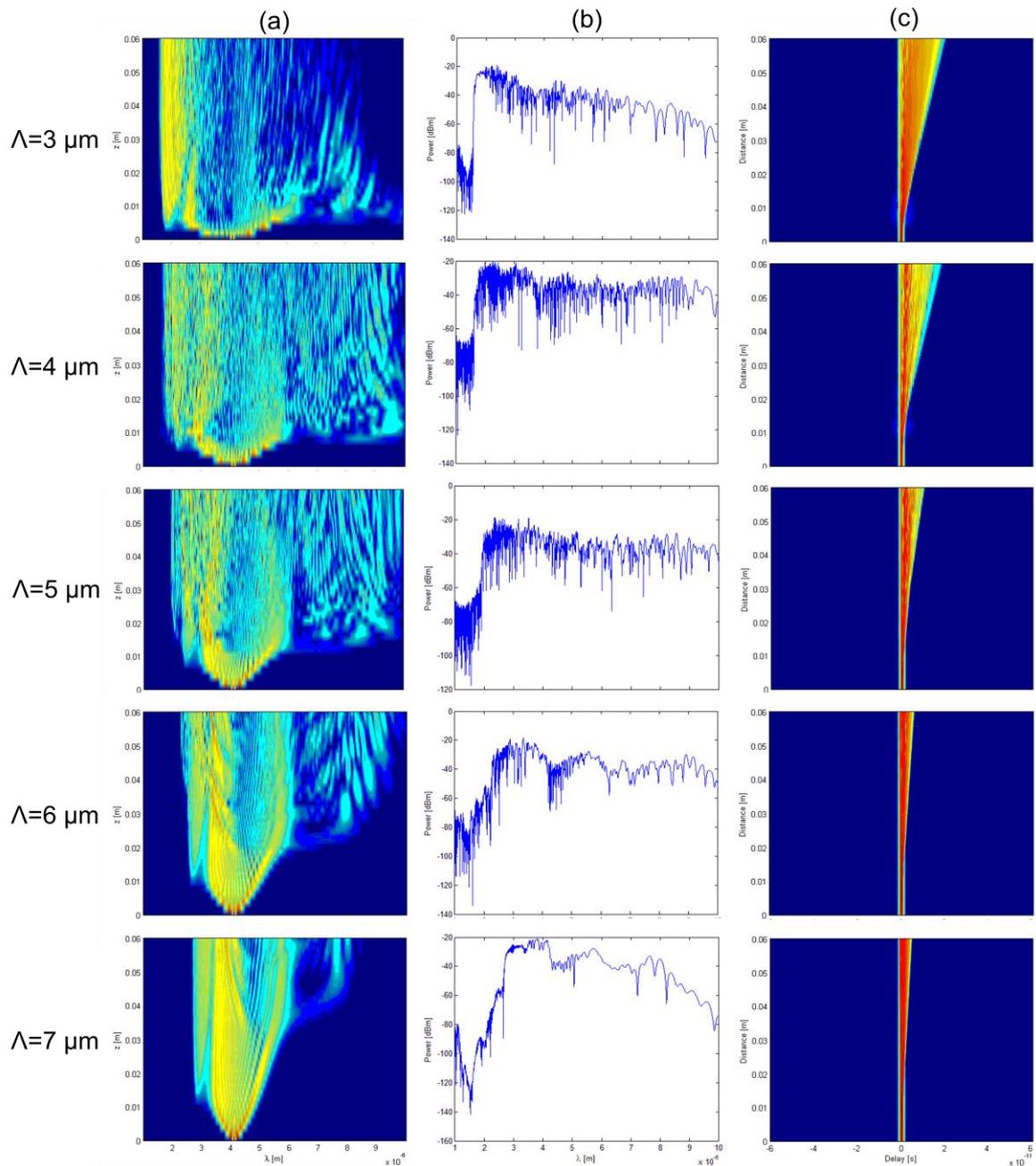

Fig.4. (a), Spectral evolutions of SC by pumping a 6 cm-long $As_2Se_3$ PCF (relative hole size d/Λ = 0.4) with 4.1 μm pulse (500 fs pulse and 10 kW peak power), (b) spectra of SC after propagation of 6 cm, (c) temporal evolutions of SC along the propagation length, for Λ = 3 μm, 4 μm, 5 μm, 6 μm, 7 μm, repectively.

With further propagation, for pumping in the anomalous regime, the increased spectral broadening in long wavelength side occurs mainly due to solitons fission and Raman soliton self-frequency shift, with the generation of the corresponding dispersive waves at the short-wavelength side of the ZDW, which eventually merge to form a broad SC spectrum. For pumping in the normalous regime, additional broadening is mainly due to the self-phase modulation (SPM) and the Raman effect.

The resulting SC spectra obtained after 6 cm propagation of the pumping pulse are shown in Fig.4 (b). For $\Lambda$= 3 μm, it can be seen that the broadening process of the SC spectrum at the long wavelength side is severly narrowed by the confiment loss and the second ZDW of PCF at long-wavelength side, which leads to an unflat spectral profile. For $\Lambda$= 6, and 7 μm, viewing from the SC spectra, it is found that the first ZDW limts the spectal expansion at the long wavelength side, in particular, for pitch of 7 μm. Only a small portion of the pump power leaks from normal dispersion regime to abnormal dispersion regime through the high-order dispersion FWM at the first few millimeters' propagation. As the first ZDW shifting to the longer wavelength and far from the pump wavelength, a larger group velocity mismatch manifests itself to an increase of the frequency seperation and a decrease in papametirc bandwidth of the FWM sidebands. Since the nonlinear coefficient $\bar{\gamma}(\omega)$ is inversely proportional to the effective area $A_{eff}$, and $A_{eff}$ increases with $\Lambda$, all other parameters including $n_{eff}$ remain largely constant, it follows that the maximum FWM gain decreases with the increase of $\Lambda$. All these effects lead to a decrease of the efficiency of the FWM process and a distinct tendency of a narrow and irregular SC generation spectrum as seen in Fig. 4 for $\Lambda$= 6 and 7 μm. However, in the case of $\Lambda$= 5 μm, the pump wavelength is very near to the first ZDW and increasing the propagation distance eventually leads to the formation of a broadband SC,similar with that of $\Lambda$= 4 μm.

At the short wavelength side, it is seen that the high material loss has a significant negative impact on the resulting SC bandwidth when wavelenght is lower than 2 μm for all study cases. It is also found from Fig. 4. (c) that, due to the flat dispersion properties, the group volocity walk-off effect of temporal pulses which leads to the development of distinct temporal peaks during the propagation does not occur . However, it is observed that the temporal delay becomes apparent when spectral broadening extends more to long wavelength side, such as for $\Lambda$= 4 μm. This is a nature temporal disfocusing process due to the accumulated abnormal dispersion by the long wavelengths and and can be avoided by using a shorter fiber, such as ~2 cm for $\Lambda$ =4 μm.

From Fig.4, it is found that, despite the high material losses above 9 μm, the spectra are very broad for $\Lambda$ =4 and 5 μm, with the broadest spectrum yielding a 20 dB bandwidth of >7 μm. We also find that, when pumping in the anomalous region and near the ZDW, the spectra are remarkably flat after the propagation of 6 cm, although it takes a longer propagation length for $\Lambda$ = 5 μm to achieve this resemblance as $\Lambda$ =4 μm. We contribute this broad mid-infrared SC generation to the appropriate pump scheme of using a subharmonic generation source of the mode-locked thulium fiber laser at ~4.1 μm, the flat dispersion profile and the endlessly single-mode property of $As_2Se_3$ PCF, combining with the high nonlinearity and the wide mid-infrared transparent window of $As_2Se_3$ chalcogenide glass, which are all essential for the SC spectral broadening process.

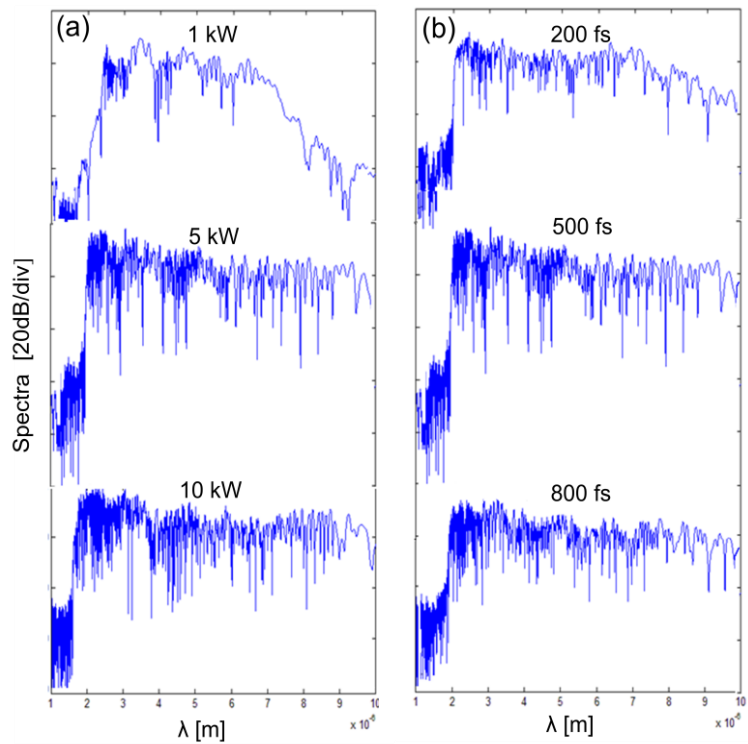

Fig.5. (a), the SC spectra of a 6 cm $As_2Se_3$ PCF ($\Lambda$ =4 and relative hole size d/$\Lambda$ = 0.4) pumped by a 4.1 μm pulse with a 500 fs pulse duration and a peak power of 1 kW, 5 kW, and 10 kW, respetively, (b), the SC spectra when a 6 cm $As_2Se_3$ PCF pumped with a 4.1 μm pulse of 5 kW and a pulse duration of 200 fs, 500 fs, and 800 fs, respectively.

To optimize the bandwidth of the SC generation, we further investigate how the pulse properties of the pump laser affects the bandwidth for a $As_2Se_3$ PCF with a pitch of 4 μm by varing the pulse peak power and the pulse duration. The simulation results illustrated in Fig.5 suggest that, for a peak power >5 kW and a sub-pecosecond pulse duration, a similar broadband MIR SC generation with the bandwidth >7 μm can be achieved. This results is expected since, as a dominative process at the beginning of the SC generation in the femtosecond regime, the efficiency of the FWM process, i.e. its gain and bandwidth, depends on both $\bar{\gamma}(\omega)$ and peak power [22]. When the $\bar{\gamma}(\omega)$ is constant, the peak power will decide the FWM gain and the initial FWM process, which lays down the fundation for the following spectral broadening due to the soliton process in the SC generation. It should be noted that the resemble final spctrum of the SC generation can be achieved for a peak power even lower than 5 kw with a longer propagation distance, however, a long fiber length is not prefered due to the high material loss.

## 4. Conclusions

In conclusion, for the first time, we demonstrate that the realization of a >7 μm broadband SC generation with a high spectral flatness is possible in a short chalcogenide PCF with a 4.1 μm pump. The maximum bandwidth occurs when pumping near to the first ZDW, such as when the pitch is 4 μm, and the 20 dB bandwidth reaches its maximum after the propagation of only a few cm fiber length. Using a subharmonic generation source of the mode-locked thulium fiber laser at ~4.1 μm is a feasible solution for achieving this broadband mid-infrared SC generation since it leverages on the existing laser sources and the chalcogenide PCF technologies. This SC scheme mitigates the troublesome post-processing procedure for chalcogenide fiber such as tapering to move the ZDW to short wavelength. As a highly potential MIR source for frequency metrology, it is critical for the resulting SC spectrum to have high coherence [12, 23]. The coherence properties of the SC generation with the pump and fiber parameters used in our simulations will be further explored in the near future.